\documentclass[aps,pra,twocolumn,groupedaddress]{revtex4-1}

\usepackage{graphicx}
\usepackage{dcolumn}
\usepackage{bm}
\usepackage{amsthm}
\usepackage{multirow}

\newtheorem*{theorem}{Theorem}

\begin{document}

\title{Non-Uniform Code Concatenation for Universal Fault-Tolerant Quantum Computing}


\author{Eesa Nikahd}
\email{nikahd@aut.ac.ir}
\author{Mehdi Sedighi}
\email{msedighi@aut.ac.ir}
\author{Morteza Saheb Zamani}
\email{szamani@aut.ac.ir}
\affiliation{Quantum Design Automation Lab \linebreak Amirkabir University of Technology, Tehran, Iran}


\date{\today}


\begin{abstract}
Using transversal gates is a straightforward and efficient technique for fault-tolerant quantum computing. Since transversal gates alone cannot be computationally universal, they must be combined with other approaches such as magic state distillation, code switching or code concatenation in order to achieve universality. In this paper we propose an alternative approach for universal fault-tolerant quantum computing mainly based on the code concatenation approach proposed in [PRL 112, 010505 (2014)] but in a non-uniform fashion. The proposed approach is described based on non-uniform concatenation of the 7-qubit Steane code with the 15-qubit Reed-Muller code as well as the 5-qubit code with the 15-qubit Reed-Muller code, which lead to two 49-qubit and 47-qubit codes, respectively. These codes can correct any arbitrary single physical error with the ability to perform a universal set of fault-tolerant gates, without using magic state distillation.
\end{abstract}


\maketitle


\section{\label{sec:intro}Introduction}
Quantum computers harness physical phenomena unique to quantum mechanics to realize a fundamentally new mode of information processing \cite{nielsen2010quantum1}. They can overcome the limitations of classical computers in efficiently solving hard computational problems for some tasks such as integer factorization \cite{shor1994algorithms2} and database search \cite{grover1996fast3}.

Unfortunately, quantum computers are highly susceptible to noise due to decoherence and imperfect quantum operations that lead to the decay of quantum information \cite{nielsen2010quantum1}\cite{unruh1995maintaining4}. Unless we can successfully mitigate the noise problem, maintaining large and coherent quantum states for a long enough time to perform quantum algorithms will not be readily possible. Quantum error correction codes have been developed to address this problem \cite{shor1995scheme5}\cite{steane1996error6}\cite{lidar2013quantum7}. To do so, data are encoded into a code and gates are applied directly on the encoded quantum states without a need to decode the states \cite{nielsen2010quantum1}. The encoded gates are applied fault-tolerantly in a way that they do not propagate errors in the circuit. Furthermore, quantum codes can be concatenated recursively to increase their ability to correct errors even further. In this way, almost perfectly reliable quantum computation is possible with noisy physical devices as long as the noise level is below a threshold value \cite{knill1996accuracy8}.

A straightforward and efficient technique for fault-tolerant quantum computing is using transversal gates. An encoded gate which can be implemented in a bitwise fashion is known as a transversal gate \cite{nielsen2010quantum1}. No quantum code with a universal set of transversal gates exists \cite{eastin2009restrictions9}. So a common solution for applying non-transversal gates is using a special state prepared by magic state distillation (MSD) protocol \cite{bravyi2005universal10}. However, the overhead of state preparation using MSD remains one of the drawbacks of this approach \cite{fowler2012surface11}. The distillation overhead scales as $O(\log_{1/\epsilon}^{\gamma})$, where $\gamma$ is determined by distillation protocol and $\epsilon$ is the desired output accuracy \cite{bravyi2012magic12}. There have been several efforts to reduce the overhead of this scheme such as \cite{bravyi2012magic12}, \cite{jones2013multilevel13} and \cite{campbell2012magic14}.

A work on universal fault-tolerant quantum computing without MSD using only one quantum error correction code has been proposed by Paetznick and Reichardt \cite{paetznick2013universal19}. In this approach, all of the gates from the considered universal set, e.g. $\{$Pauli gates, $H, CCZ\}$ have been implemented transversally, where $H$ and $CCZ$ are Hadamard and controlled-controlled-Z, respectively. However, as applying transversal $H$ gate disturbs the code space, additional error correction and transversal measurements are needed in order to recover the code space after application of this gate.

Recently, similar approaches for universal quantum computing without using MSD have been proposed. These approaches are based on combining two different codes, say $C_1$ and $C_2$, where each non-transversal gate in $C_1$ has a transversal implementation on $C_2$ and vice versa. This approach is pursued in two different ways: (1) by combining $C_1$ and $C_2$ based on code switching \cite{stephens2008asymmetric15}\cite{anderson2014fault16}\cite{choi2015dual17} and (2) by combining $C_1$ and $C_2$ in a uniform concatenated fashion \cite{jochym2014using18}. We call a concatenated code \emph{uniform}, if it uses only one quantum code in each level to encode all of the qubits of that level.

In the code switching scheme, since the two selected codes have different sets of transversal gates, one can implement a universal set of gates transversally, by switching to $C_2$ for transversal implementation of a gate which is non-transversal in $C_1$. However, a fault-tolerant switching circuit is needed which imposes an additional cost and thus, in some cases it may incur a higher cost compared to MSD \cite{choi2015dual17}. On the other hand, in the uniform code concatenation method, the logical information is encoded by $C_1$ where each qubit of $C_1$ is in turn encoded into the code of $C_2$. Therefore, the number of necessary physical qubits to code the logical information is relatively large (the product of the number of qubits for the two codes). For instance, if 7-qubit Steane and 15-qubit Reed-Muller codes are used, a code [[105, 1, 9]] will be produced. However, the code has the ability to correct only one arbitrary single error because of error propagation in a codeword during $T$ and $H$ implementation, where $T=diag(1, exp^{\frac{i\pi}{4}})$.

Recently, Yoder at el. \cite{yoder2016universal25} proposed the pieceable fault-tolerant concept to provide universal fault-tolerance by developing non-transversal, yet still fault-tolerant gates. In this approach a non-transversal circuit is broken into fault-tolerant pieces and rounds of intermediate error correction is applied in between to correct errors before they become uncorrectable by propagating in the circuit.

In this paper, we propose a new method for universal fault-tolerant quantum computing mainly based on code concatenation approach proposed in \cite{jochym2014using18}, but in a non-uniform fashion. The proposed method opens a new perspective of code concatenation for universal fault-tolerant computation by considering the structural details of non-transversal gates and reduces the overhead of the uniform code concatenation method proposed in \cite{jochym2014using18}.

\section{\label{sec:proposed}Non-uniform code concatenation}
\begin{figure}
	\centering
	\includegraphics[trim=0in 0in 0in 0in,clip,width=0.9\columnwidth]{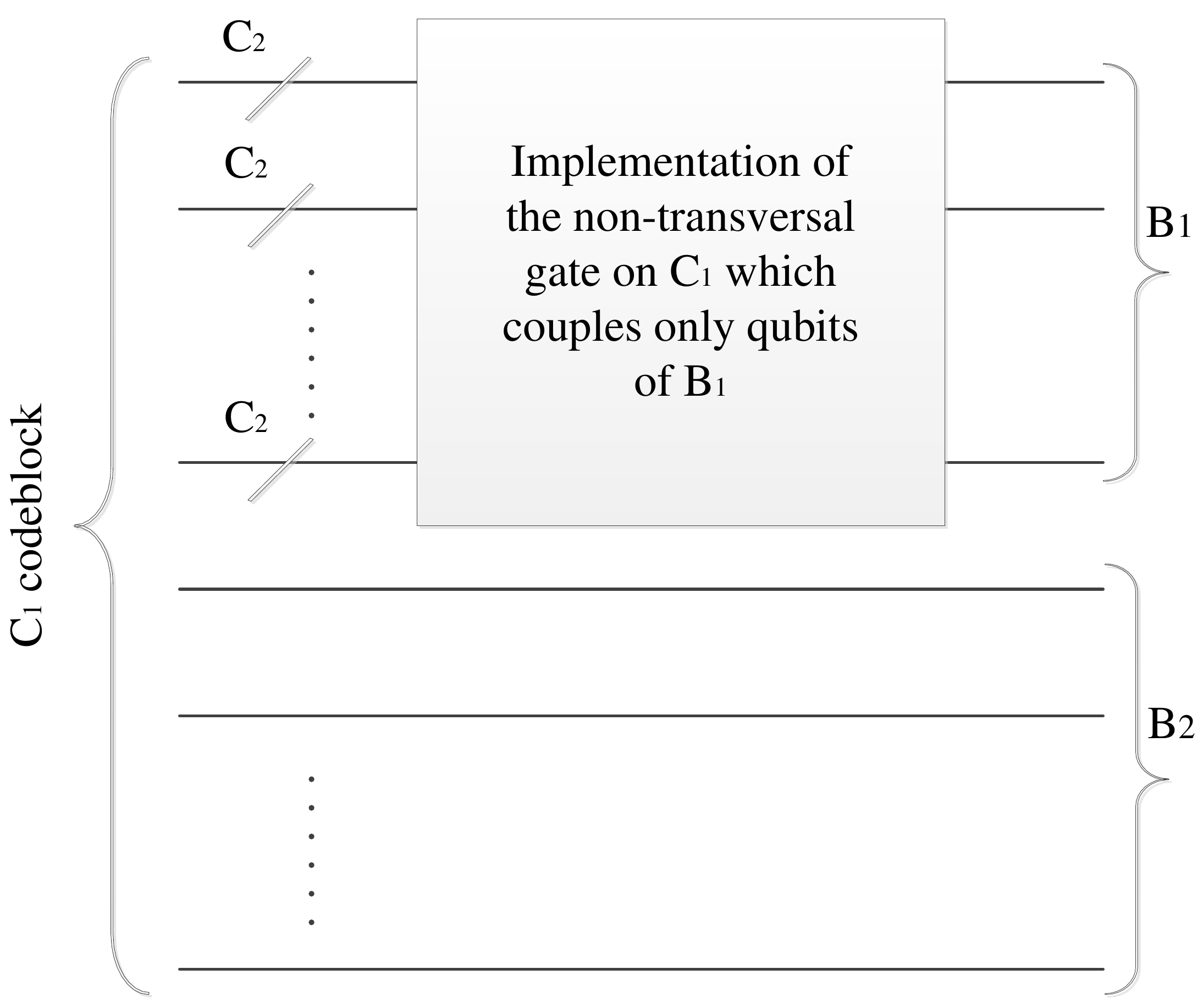}
	\caption{The schematic overview of the non-uniform code concatenation approach. Logical information is encoded by $C_1$. In the second level of concatenation each qubit of $B_1$ is in turn encoded into the code of $C_2$ and the $B_2$ qubits are left unchanged without encoding.}
	\label{fig:overall}
\end{figure}

The proposed approach is based on non-uniform code concatenation of $C_1$ and $C_2$. In this approach, a logical qubit is encoded using $C_1$ in the first level of coding hierarchy. However, in the second level, only some of the $C_1$ qubits are encoded using $C_2$ depending on the implementation of non-transversal gates in $C_1$, as opposed to \cite{jochym2014using18} which encodes all of the $C_1$ qubits using $C_2$ in the second level of concatenation. The remaining qubits can be encoded using $C_1$ or remain unchanged. In contrast with uniform concatenated codes, we call such a code \emph{non-uniform}, which uses more than one code in at least one level of its coding hierarchy. The idea of non-uniform code concatenation is motivated by the observation that application of a non-transversal gate in $C_1$ does not necessarily involve all of the $C_1$ qubits. Therefore, it is not necessary to encode all of the $C_1$ qubits using $C_2$. The $C_1$ qubits can be partitioned into two non-overlapping sets: the set $B_1$ which contains qubits that are coupled during the application of non-transversal gate in $C_1$ and $B_2$ which contains the uncoupled qubits. Indeed, we only need to encode qubits of $B_1$ using $C_2$ and can leave the $B_2$ qubits unchanged. If there is more than one non-transversal gate in $C_1$, the set $B_1$ contains the union of all involved qubits in the implementation of each non-transversal gate. Fig. \ref{fig:overall} depicts a schematic overview of the proposed approach.

$C_1$ and $C_2$ must have the properties described in \cite{jochym2014using18}. 1) $C_1$ and $C_2$ must have at least a distance of three. 2) For any logical gate in the universal gate set with non-transversal implementation on $C_1$, there must exist an equivalent transversal implementation on $C_2$. 3) The error correction operations and syndrome measurement on $C_1$ and $C_2$ must be globally transversal in the concatenated code space. However, for our method to produce superior results compared to \cite{jochym2014using18}, it is necessary to have $\left| B_2 \right| >0$. Fortunately, for a stabilizer code, there is a useful family of gates which can be implemented by coupling only $d$ qubits, where $d$ is the code distance.

\begin{figure}
	\centering
	\includegraphics[trim=0in 0in 0in 0in,clip,width=0.9\columnwidth]{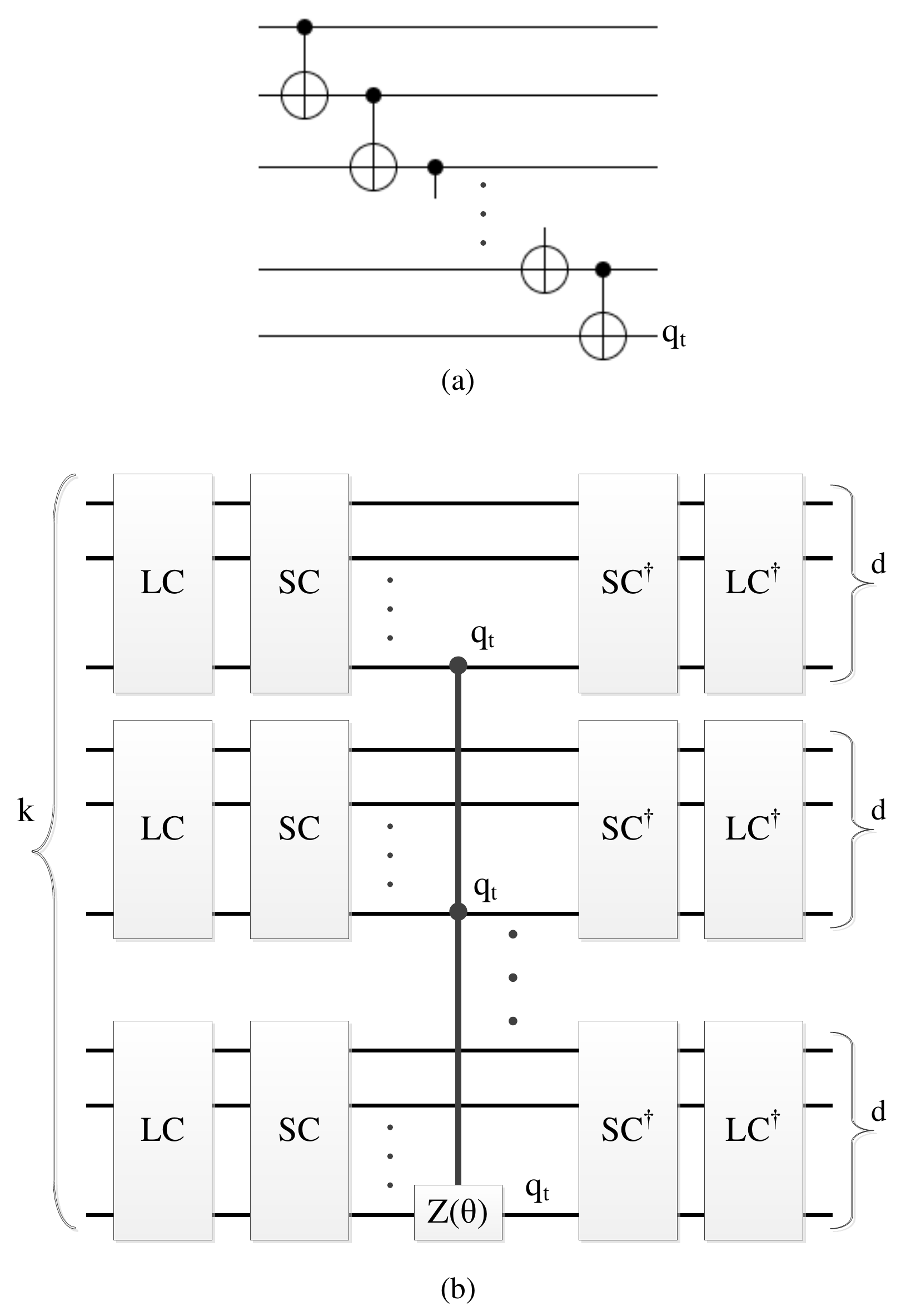}
	\caption{(a) Staircase of $CNOT$s. (b) Non-transversal application of $C^{k}Z(\theta)$ gate for a stabilizer code by involving only $d$ qubits of each code block, where $d$ is the code distance. $SC$ is an acronym for staircase of $CNOT$s and $LC$ is a circuit containing only local Clifford gates which transform the original logical $Z$ operator into a form consisting of only Pauli $Z$’s and $I$’s. Note that only the qubits of $B_1$ are shown.}
	\label{fig:CkZ}
\end{figure}

\begin{theorem}
	For every stabilizer code, a logical $C^{k}Z(\theta)$ gate can be implemented non-transversally by coupling only $d$ qubits, where $d$ is the code distance and $Z(\theta)=diag(1,exp^{i\theta})$.
\end{theorem}
\begin{proof}
	When the code distance is $d$, there is a Pauli operator in the normalizer of the stabilizer group (that does not belong to the stabilizer group itself) with weight $d$, where weight of an $n$-qubit Pauli operator is defined as the number of its non-identity members. This operator is a logical operator and can be thought as a logical $Z$ gate. This operator can be transformed into a form consisting of only Pauli $Z$’s and $I$’s with positive sign by applying local Clifford gates. The ability to perform a logical $Z$ by applying $d$ physical $Z$ gates on $d$ distinct qubits means that an even (odd) number of these $d$ qubits are $|1\rangle$ in each term of the logical qubit state, where the logical qubit is in the state $|\overline{0}\rangle$ ($|\overline{1}\rangle$). Thus, applying a staircase of $CNOT$ gates on these $d$ qubits (as shown in Fig. \ref{fig:CkZ} $(a)$) leaves the last target physical qubit ($q_t$) into the physical state $|\psi \rangle$, where the logical qubit is in the state $|\overline{\psi}\rangle$. Therefore, applying a single physical gate $Z(\theta)$ on $q_t$ acts exactly as logical one. Similarly, a logical $C^{k}Z(\theta)$ can also be implemented using a single physical $C^{k}Z(\theta)$ on $q_t$s of $k$ logical qubits (Fig. \ref{fig:CkZ} $(b)$).	
\end{proof}

Fault-tolerant application of non-transversal gates in $C_1$, non-transversal gates in $C_2$ and error correction procedure in the proposed code are described in the following.%

\subsection{Fault-tolerant implementation of the non-transversal gates in $C_1$}
A single physical error on one of the qubits of $B_1$, occurring in the non-transversal application of these gates on $C_1$ only propagates between the qubits of $B_1$, which are themselves encoded blocks of $C_2$. Since implementations of these gates on $C_1$ consist of only transversal gates in $C_2$, this single physical error only propagates to a single physical error in each of the $B_1$ qubits. As these qubits are encoded using $C_2$, this single physical error can be corrected by error correction procedure on $C_2$ code blocks.

A single physical error on the $B_2$ qubits during application of these non-transversal gates in $C_1$ does not propagate to any other qubits of $C_1$ code block and can be corrected using the error correction procedure on $C_1$.

\subsection{Fault-tolerant implementation of the non-transversal gates in $C_2$}
These gates have transversal implementation on $C_1$ and therefore, a single physical error on one of the $C_1$ qubits, does not propagate to any other qubits of $C_1$, during application of these gates. But as they are non-transversal in $C_2$, a single error on a particular $C_2$ code block (qubits of $B_1$) can propagate to a non-correctable set of errors on that code block which introduces a $C_2$ logical error. However, this error only leads to a single error on one of the qubits of $C_1$ which can be corrected using the error correction procedure on $C_1$.

\subsection{Error correction procedure}
Regarding the third necessary condition for code concatenation, error correction procedure are globally transversal and therefore, fault-tolerant in the concatenated code space. This feature is essential not only for preventing error propagation during the error correction procedure, as described in \cite{jochym2014using18}, it also makes the non-uniform code concatenation possible. Indeed, this feature guarantees that there is no interaction among qubits of the sets $B_1$ and $B_2$ which are encoded blocks of different codes, during error correction procedure. How the error correction should be applied for non-uniform concatenated codes is the same as uniform one. 

Although straight concatenation of the two codes [[$n_1$,$k$,$d_1$]] and [[$n_2$,1,$d_2$]] leads to a code [[$n_1$$n_2$,$k$,$d_1$$d_2$]] \cite{gottesman1997stabilizer22}, our code concatenation scheme reduces the effective distance of the concatenated code in order to achieve universal fault tolerance. By effective distance we mean the code distance considering the error propagations that occur during application of the non-transversal gates. 

While the proposed approach is general and can be applied to any code combination that satisfies the mentioned conditions, in the rest of paper we will focus on the 7-qubit Steane and 5-qubit quantum error correction codes (the smallest quantum codes with distance of three), as $C_1$, in combination with 15-qubit Reed-Muller (RM) code (the smallest known quantum code with transversal $T$ and $CCZ$ gate), as $C_2$.

\subsection{Non-uniform concatenation of the Steane and 15-qubit Reed-Muller codes}
\begin{figure}
	\centering
	\includegraphics[trim=0in 0in 0in 0in,clip,width=0.9\columnwidth]{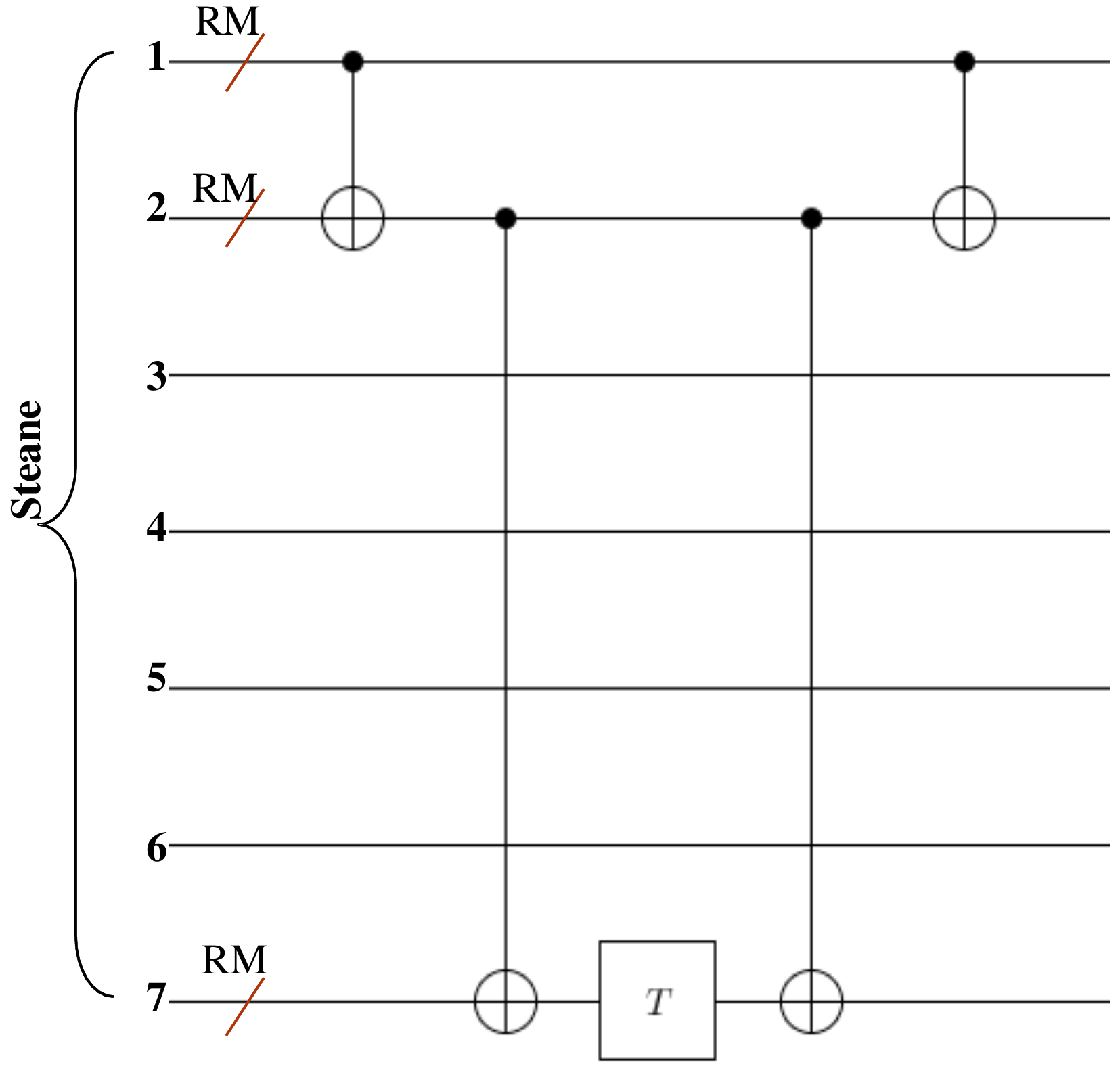}
	\caption{Fault-tolerant application of the $T$ gate for the proposed 49-qubit non-uniform concatenated code. A logical qubit is encoded by Steane where qubits 1, 2, and 7 are themselves encoded blocks of RM code and the other four qubits are left unchanged.}
	\label{fig:49qubits}
\end{figure}

The universal set \{$H$, $S$, $T$, $CNOT$\} is chosen as the gate library in this section, where $S=T^2$. $S$, $H$ and $CNOT$ and therefore, any gates from the Clifford group have transversal implementation on the Steane code. The $T$ gate remains the only non-transversal gate from the set. As shown in Fig. \ref{fig:49qubits}, for this code $B_1=\{1,2,7\}$ and $B_2= \{3,4,5,6\}$. The $T$ gate is transversal in the RM code but the Hadammard gate is not \cite{jochym2014using18}. Both of these codes have distance of three and the combination of their set of transversal gates produces a universal gate set. Based on the non-uniform approach, there is no need to encode all of the Steane qubits using the RM code. We need only to encode qubits of $B_1$ using the RM code and can leave the $B_2$ qubits unencoded. Doing so, a 49-qubit code is constructed which has the ability to correct any single physical error like the 105-qubit code proposed in \cite{jochym2014using18} but with substantial improvement in resource overhead as the number of qubits and operations are reduced significantly.

As both the Steane and RM quantum codes have the same property that $S$ and $CNOT$ gates can be implemented transversally, then for the proposed 49-qubit code they have also transversal implementation. Additionally, all syndrome measurements and Pauli corrections will be transversal within both codes \cite{jochym2014using18} and therefore, error correction procedure on the Steane and RM code blocks are globally transversal and fault-tolerant in the 49-qubit code space. 

The $CCZ=C^{2}Z(\pi)$ can also be applied fault-tolerantly for the proposed 49-qubit code, as its implementation on the Steane code has the same structure as $T$ and it is transversal in the RM code.

\subsection{Concatenation of the 5-qubit code with the 15-qubit Reed-Muller code}
\begin{figure}
	\centering
	\includegraphics[trim=0in 0in 0in 0in,clip,width=0.9\columnwidth]{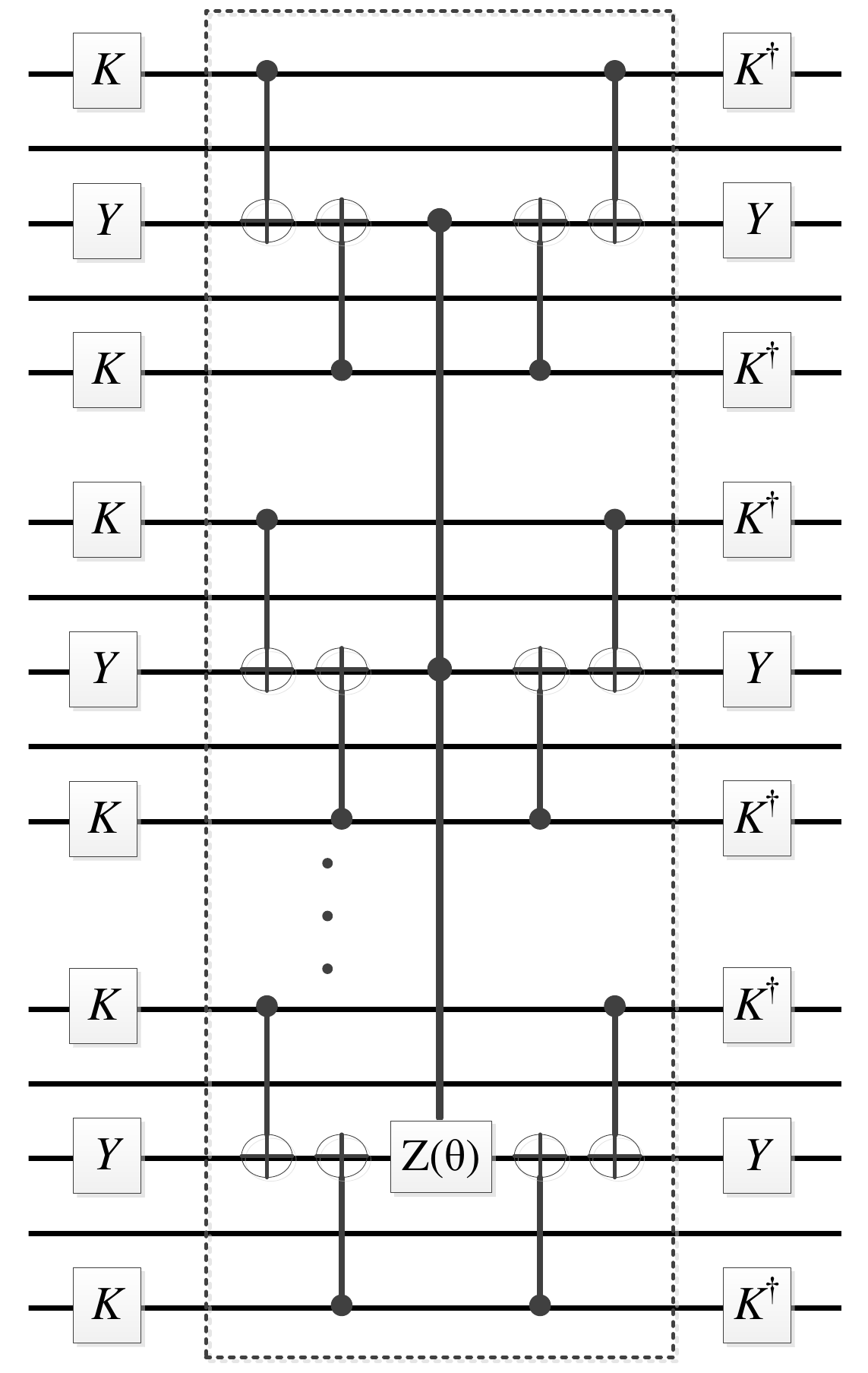}
	\caption{Non-transversal implementation of the $C^{k}Z(\theta)$ gate for the 5-qubit code. The circuit specified by the dotted box shows the implementation of this gate for the 5'-qubit code.}
	\label{fig:CKZ5}
\end{figure}

Let M=\{$T=C^{0}Z(\frac{\pi}{4})$, $S=C^{0}Z(\frac{\pi}{2})$, $CZ=C^{1}Z(\pi)$, $CCZ=C^{2}Z(\pi)$\}. The gates of $M$ along with $H$ provide a universal set for quantum computation. $H$ is transversal for the 5-qubit code but by permutation \cite{yoder2016universal25}. However, this permutation is in contrast with the nature of non-uniformity which makes it unusable for our non-uniform construction. The gates of $M$ (generally $C^{k}Z(\theta)$) can be applied non-transversally on the 5-qubit code block as shown in Fig. \ref{fig:CKZ5}, where $K=SH$. Note that $K$ is not transversal in the RM code. Therefore, non-transversal implementation of the gates of $M$ on the 5-qubit code does not involve only gates that are transversal in RM and thus, violates the second necessary condition for code concatenation. 

Therefore, the 5-qubit code in the standard form does not satisfy the needed condition for non-uniform code concatenation. However, one can alter this code to an equivalent code, namely 5'-qubit code, by applying $K_1Y_3K_5$ on the 5-qubit code block \cite{yoder2016universal25}. For this code, the $K$ gate can be applied transversally as $Z_3K^{\otimes5}$. The gates of $M$ can also be implemented as shown in the dotted box of Fig. \ref{fig:CKZ5}. This implementation only consists of the gates that are transversal in the RM code and therefore satisfies the second condition for code concatenation.

As $H=S^\dagger K$, $K$ along with the gates of $M$ provide a universal set of quantum gates. This set is considered as the gate library for the codes proposed in this section. Considering the 5'-qubit code as $C_1$ and the RM code as $C_2$ satisfies the necessary condition for code concatenation regarding this universal set. The concatenation of these codes uniformly leads to a 75-qubit code where all of the $C_1$ qubits are encoded blocks of RM. Furthermore, for the 5'-qubit code $B_1=\{1,3,5\}$ and $B_2=\{2,4\}$. Therefore, non-uniform concatenation of them produces a 47-qubit code where the $B_1$ qubits are encoded by the RM code in the second level of concatenation and the qubits of $B_2$ are left unencoded. Both the 75-qubit and 47-qubit codes have the ability to perform the gates of universal set, fault-tolerantly. 

\section{\label{sec:discussion}Discussion}

It is worth mentioning that the proposed non-uniform 47 and 49-qubit codes reduce the overall distance of their corresponding uniform codes (e.g. 75 and 105-qubit codes, respectively), as they leave the qubits of $B_2$ unencoded. Nevertheless, the $B_2$ qubits can be encoded using the $C_1$ code in the second level of concatenation which leads to two 55 and 73-qubit codes, respectively. Doing so will increase the overall distance of the codes to 9 like the uniform ones. However, in the worst case, the effective distance of these codes remains unchanged with the ability to correct a single physical error. This is because two physical errors on the qubits of $B_1$ may corrupt all of the $B_1$ qubits during application of the non-transversal gates which cannot be corrected using $C_1$ error correction procedure and therefore, leads to a logical error. Table \ref{tab:comparison} compares the produced concatenated codes based on the Steane and RM codes in terms of number of qubits, overall distance and effective distance.

The 7-qubit Steane and 15-qubit Reed-Muller (RM) codes have unique features as follows which make their concatenation efficient. The Steane code is the smallest CSS code with distance of three and with the ability to implement a universal set of Clifford gates, transversally. The $T$ gate is a non-transversal gate in the Steane code which can be applied by involving only three qubits (Fig. \ref{fig:49qubits}) and along with the Clifford gates provides a universal set of gates. The RM code is the smallest known code with transversal $T$ gate and also a CSS code. Therefore, their concatenation leads to the smallest concatenated CSS code based on the proposed approach with the ability to perform a universal set of fault-tolerant gates. It should be noted that the CSS codes have some useful properties which make them good choices for fault-tolerant quantum computation \cite{gottesman2006quantum24}. Furthermore, the Steane and RM codes have the minimum number of unshared transversal gates, e.g. $T$ and $H$.
While the codes produced using the 5'-qubit code have fewer qubits, they are non-CSS and also the effective distance of the concatenated codes is reduced for all of the gates from universal set. This is because there are no shared transversal gates between the 5'-qubit and RM codes. 


\begin{table}%
	\begin{center}
		\caption{Comparison of the produced concatenated codes based on the Steane and RM codes in terms of number of qubits, overall distance and effective distance.}
		{\footnotesize
			\begin{tabular}{|l|l|l|l|}
				\hline
				\multirow{2}{1.2in}{code concatenation \\ method} &	\# qubits	& \multirow{2}{0.5in}{overall \\ distance} 	& \multirow{2}{0.5in}{effective \\ distance}\\
				&&&\\\hline
				uniform \cite{jochym2014using18}	& $105$	& $9$	& $3$\\\hline
				non-uniform & $49$	& $5$ \cite{chamberland2016architectural26}	& $3$\\\hline
				non-uniform	& $75$	& $9$	& $3$\\\hline
			\end{tabular}
		}
		\label{tab:comparison}
	\end{center}
\end{table}

\section{\label{sec:conclusion}Conclusion}
In this paper, a non-uniform code concatenation approach was proposed for fault-tolerant quantum computing without using MSD. Four 47, 49, 55 and 73-qubit codes are constructed based on this approach with the ability to correct an arbitrary single physical error which outperforms their counterpart uniform concatenated codes. Introducing the non-uniform code concatenation concept and exploiting it in design of a new universal fault-tolerant quantum computation method by considering the implementation details of the non-transversal gates in $C_1$, is the main contribution of the proposed approach. It is worth noting that in such code concatenation schemes (both uniform and non-uniform) the effective distance of the concatenated code is reduced in order to make the universal fault-tolerant computation possible. Although, the proposed approach was described based on the 5-qubit and Steane codes in concatenation with the 15-qubit Reed-Muller code, one may pursue this work by investigating other code combinations.

\bibliography{NUCC_UFQC}

\providecommand{\noopsort}[1]{}\providecommand{\singleletter}[1]{#1}%
\begin{thebibliography}{23}%
\makeatletter
\providecommand \@ifxundefined [1]{%
 \@ifx{#1\undefined}
}%
\providecommand \@ifnum [1]{%
 \ifnum #1\expandafter \@firstoftwo
 \else \expandafter \@secondoftwo
 \fi
}%
\providecommand \@ifx [1]{%
 \ifx #1\expandafter \@firstoftwo
 \else \expandafter \@secondoftwo
 \fi
}%
\providecommand \natexlab [1]{#1}%
\providecommand \enquote  [1]{``#1''}%
\providecommand \bibnamefont  [1]{#1}%
\providecommand \bibfnamefont [1]{#1}%
\providecommand \citenamefont [1]{#1}%
\providecommand \href@noop [0]{\@secondoftwo}%
\providecommand \href [0]{\begingroup \@sanitize@url \@href}%
\providecommand \@href[1]{\@@startlink{#1}\@@href}%
\providecommand \@@href[1]{\endgroup#1\@@endlink}%
\providecommand \@sanitize@url [0]{\catcode `\\12\catcode `\$12\catcode
  `\&12\catcode `\#12\catcode `\^12\catcode `\_12\catcode `\%12\relax}%
\providecommand \@@startlink[1]{}%
\providecommand \@@endlink[0]{}%
\providecommand \url  [0]{\begingroup\@sanitize@url \@url }%
\providecommand \@url [1]{\endgroup\@href {#1}{\urlprefix }}%
\providecommand \urlprefix  [0]{URL }%
\providecommand \Eprint [0]{\href }%
\providecommand \doibase [0]{http://dx.doi.org/}%
\providecommand \selectlanguage [0]{\@gobble}%
\providecommand \bibinfo  [0]{\@secondoftwo}%
\providecommand \bibfield  [0]{\@secondoftwo}%
\providecommand \translation [1]{[#1]}%
\providecommand \BibitemOpen [0]{}%
\providecommand \bibitemStop [0]{}%
\providecommand \bibitemNoStop [0]{.\EOS\space}%
\providecommand \EOS [0]{\spacefactor3000\relax}%
\providecommand \BibitemShut  [1]{\csname bibitem#1\endcsname}%
\let\auto@bib@innerbib\@empty
\bibitem [{\citenamefont {Nielsen}\ and\ \citenamefont
  {Chuang}(2010)}]{nielsen2010quantum1}%
  \BibitemOpen
  \bibfield  {author} {\bibinfo {author} {\bibfnamefont {M.~A.}\ \bibnamefont
  {Nielsen}}\ and\ \bibinfo {author} {\bibfnamefont {I.~L.}\ \bibnamefont
  {Chuang}},\ }\href@noop {} {\emph {\bibinfo {title} {Quantum computation and
  quantum information}}}\ (\bibinfo  {publisher} {Cambridge university press},\
  \bibinfo {year} {2010})\BibitemShut {NoStop}%
\bibitem [{\citenamefont {Shor}(1994)}]{shor1994algorithms2}%
  \BibitemOpen
  \bibfield  {author} {\bibinfo {author} {\bibfnamefont {P.~W.}\ \bibnamefont
  {Shor}},\ }in\ \href@noop {} {\emph {\bibinfo {booktitle} {Foundations of
  Computer Science, 1994 Proceedings., 35th Annual Symposium on}}}\ (\bibinfo
  {organization} {IEEE},\ \bibinfo {year} {1994})\ pp.\ \bibinfo {pages}
  {124--134}\BibitemShut {NoStop}%
\bibitem [{\citenamefont {Grover}(1996)}]{grover1996fast3}%
  \BibitemOpen
  \bibfield  {author} {\bibinfo {author} {\bibfnamefont {L.~K.}\ \bibnamefont
  {Grover}},\ }in\ \href@noop {} {\emph {\bibinfo {booktitle} {Proceedings of
  the twenty-eighth annual ACM symposium on Theory of computing}}}\ (\bibinfo
  {organization} {ACM},\ \bibinfo {year} {1996})\ pp.\ \bibinfo {pages}
  {212--219}\BibitemShut {NoStop}%
\bibitem [{\citenamefont {Unruh}(1995)}]{unruh1995maintaining4}%
  \BibitemOpen
  \bibfield  {author} {\bibinfo {author} {\bibfnamefont {W.~G.}\ \bibnamefont
  {Unruh}},\ }\href@noop {} {\bibfield  {journal} {\bibinfo  {journal} {Phys.\
  Rev. A}\ }\textbf {\bibinfo {volume} {51}},\ \bibinfo {pages} {992} (\bibinfo
  {year} {1995})}\BibitemShut {NoStop}%
\bibitem [{\citenamefont {Shor}(1995)}]{shor1995scheme5}%
  \BibitemOpen
  \bibfield  {author} {\bibinfo {author} {\bibfnamefont {P.~W.}\ \bibnamefont
  {Shor}},\ }\href@noop {} {\bibfield  {journal} {\bibinfo  {journal} {Phys.
  Rev. A}\ }\textbf {\bibinfo {volume} {52}},\ \bibinfo {pages} {R2493}
  (\bibinfo {year} {1995})}\BibitemShut {NoStop}%
\bibitem [{\citenamefont {Steane}(1996)}]{steane1996error6}%
  \BibitemOpen
  \bibfield  {author} {\bibinfo {author} {\bibfnamefont {A.~M.}\ \bibnamefont
  {Steane}},\ }\href@noop {} {\bibfield  {journal} {\bibinfo  {journal}
  {Physical Review Letters}\ }\textbf {\bibinfo {volume} {77}},\ \bibinfo
  {pages} {793} (\bibinfo {year} {1996})}\BibitemShut {NoStop}%
\bibitem [{\citenamefont {Lidar}\ and\ \citenamefont
  {Brun}(2013)}]{lidar2013quantum7}%
  \BibitemOpen
  \bibfield  {author} {\bibinfo {author} {\bibfnamefont {D.~A.}\ \bibnamefont
  {Lidar}}\ and\ \bibinfo {author} {\bibfnamefont {T.~A.}\ \bibnamefont
  {Brun}},\ }\href@noop {} {\emph {\bibinfo {title} {Quantum error
  correction}}}\ (\bibinfo  {publisher} {Cambridge University Press},\ \bibinfo
  {year} {2013})\BibitemShut {NoStop}%
\bibitem [{\citenamefont {Knill}\ \emph {et~al.}()\citenamefont {Knill},
  \citenamefont {Laflamme},\ and\ \citenamefont {Zurek}}]{knill1996accuracy8}%
  \BibitemOpen
  \bibfield  {author} {\bibinfo {author} {\bibfnamefont {E.}~\bibnamefont
  {Knill}}, \bibinfo {author} {\bibfnamefont {R.}~\bibnamefont {Laflamme}}, \
  and\ \bibinfo {author} {\bibfnamefont {W.}~\bibnamefont {Zurek}},\
  }\href@noop {} {\bibinfo  {journal} {Citeseer:10.1.1.55.8881}\ }\BibitemShut
  {NoStop}%
\bibitem [{\citenamefont {Eastin}\ and\ \citenamefont
  {Knill}(2009)}]{eastin2009restrictions9}%
  \BibitemOpen
\bibfield  {journal} {  }\bibfield  {author} {\bibinfo {author} {\bibfnamefont
  {B.}~\bibnamefont {Eastin}}\ and\ \bibinfo {author} {\bibfnamefont
  {E.}~\bibnamefont {Knill}},\ }\href@noop {} {\bibfield  {journal} {\bibinfo
  {journal} {Physical review letters}\ }\textbf {\bibinfo {volume} {102}},\
  \bibinfo {pages} {110502} (\bibinfo {year} {2009})}\BibitemShut {NoStop}%
\bibitem [{\citenamefont {Bravyi}\ and\ \citenamefont
  {Kitaev}(2005)}]{bravyi2005universal10}%
  \BibitemOpen
  \bibfield  {author} {\bibinfo {author} {\bibfnamefont {S.}~\bibnamefont
  {Bravyi}}\ and\ \bibinfo {author} {\bibfnamefont {A.}~\bibnamefont
  {Kitaev}},\ }\href@noop {} {\bibfield  {journal} {\bibinfo  {journal}
  {Physical Review A}\ }\textbf {\bibinfo {volume} {71}},\ \bibinfo {pages}
  {022316} (\bibinfo {year} {2005})}\BibitemShut {NoStop}%
\bibitem [{\citenamefont {Fowler}\ \emph {et~al.}(2012)\citenamefont {Fowler},
  \citenamefont {Mariantoni}, \citenamefont {Martinis},\ and\ \citenamefont
  {Cleland}}]{fowler2012surface11}%
  \BibitemOpen
  \bibfield  {author} {\bibinfo {author} {\bibfnamefont {A.~G.}\ \bibnamefont
  {Fowler}}, \bibinfo {author} {\bibfnamefont {M.}~\bibnamefont {Mariantoni}},
  \bibinfo {author} {\bibfnamefont {J.~M.}\ \bibnamefont {Martinis}}, \ and\
  \bibinfo {author} {\bibfnamefont {A.~N.}\ \bibnamefont {Cleland}},\
  }\href@noop {} {\bibfield  {journal} {\bibinfo  {journal} {Physical Review
  A}\ }\textbf {\bibinfo {volume} {86}},\ \bibinfo {pages} {032324} (\bibinfo
  {year} {2012})}\BibitemShut {NoStop}%
\bibitem [{\citenamefont {Bravyi}\ and\ \citenamefont
  {Haah}(2012)}]{bravyi2012magic12}%
  \BibitemOpen
  \bibfield  {author} {\bibinfo {author} {\bibfnamefont {S.}~\bibnamefont
  {Bravyi}}\ and\ \bibinfo {author} {\bibfnamefont {J.}~\bibnamefont {Haah}},\
  }\href@noop {} {\bibfield  {journal} {\bibinfo  {journal} {Physical Review
  A}\ }\textbf {\bibinfo {volume} {86}},\ \bibinfo {pages} {052329} (\bibinfo
  {year} {2012})}\BibitemShut {NoStop}%
\bibitem [{\citenamefont {Jones}(2013)}]{jones2013multilevel13}%
  \BibitemOpen
  \bibfield  {author} {\bibinfo {author} {\bibfnamefont {C.}~\bibnamefont
  {Jones}},\ }\href@noop {} {\bibfield  {journal} {\bibinfo  {journal}
  {Physical Review A}\ }\textbf {\bibinfo {volume} {87}},\ \bibinfo {pages}
  {042305} (\bibinfo {year} {2013})}\BibitemShut {NoStop}%
\bibitem [{\citenamefont {Campbell}\ \emph {et~al.}(2012)\citenamefont
  {Campbell}, \citenamefont {Anwar},\ and\ \citenamefont
  {Browne}}]{campbell2012magic14}%
  \BibitemOpen
  \bibfield  {author} {\bibinfo {author} {\bibfnamefont {E.~T.}\ \bibnamefont
  {Campbell}}, \bibinfo {author} {\bibfnamefont {H.}~\bibnamefont {Anwar}}, \
  and\ \bibinfo {author} {\bibfnamefont {D.~E.}\ \bibnamefont {Browne}},\
  }\href@noop {} {\bibfield  {journal} {\bibinfo  {journal} {Physical Review
  X}\ }\textbf {\bibinfo {volume} {2}},\ \bibinfo {pages} {041021} (\bibinfo
  {year} {2012})}\BibitemShut {NoStop}%
\bibitem [{\citenamefont {Paetznick}\ and\ \citenamefont
  {Reichardt}(2013)}]{paetznick2013universal19}%
  \BibitemOpen
  \bibfield  {author} {\bibinfo {author} {\bibfnamefont {A.}~\bibnamefont
  {Paetznick}}\ and\ \bibinfo {author} {\bibfnamefont {B.~W.}\ \bibnamefont
  {Reichardt}},\ }\href@noop {} {\bibfield  {journal} {\bibinfo  {journal}
  {Physical review letters}\ }\textbf {\bibinfo {volume} {111}},\ \bibinfo
  {pages} {090505} (\bibinfo {year} {2013})}\BibitemShut {NoStop}%
\bibitem [{\citenamefont {Stephens}\ \emph {et~al.}(2008)\citenamefont
  {Stephens}, \citenamefont {Evans}, \citenamefont {Devitt},\ and\
  \citenamefont {Hollenberg}}]{stephens2008asymmetric15}%
  \BibitemOpen
  \bibfield  {author} {\bibinfo {author} {\bibfnamefont {A.~M.}\ \bibnamefont
  {Stephens}}, \bibinfo {author} {\bibfnamefont {Z.~W.~E.}\ \bibnamefont
  {Evans}}, \bibinfo {author} {\bibfnamefont {S.~J.}\ \bibnamefont {Devitt}}, \
  and\ \bibinfo {author} {\bibfnamefont {L.~C.~L.}\ \bibnamefont
  {Hollenberg}},\ }\href@noop {} {\bibfield  {journal} {\bibinfo  {journal}
  {Physical Review A}\ }\textbf {\bibinfo {volume} {77}},\ \bibinfo {pages}
  {062335} (\bibinfo {year} {2008})}\BibitemShut {NoStop}%
\bibitem [{\citenamefont {Anderson}\ \emph {et~al.}(2014)\citenamefont
  {Anderson}, \citenamefont {Duclos-Cianci},\ and\ \citenamefont
  {Poulin}}]{anderson2014fault16}%
  \BibitemOpen
  \bibfield  {author} {\bibinfo {author} {\bibfnamefont {J.~T.}\ \bibnamefont
  {Anderson}}, \bibinfo {author} {\bibfnamefont {G.}~\bibnamefont
  {Duclos-Cianci}}, \ and\ \bibinfo {author} {\bibfnamefont {D.}~\bibnamefont
  {Poulin}},\ }\href@noop {} {\bibfield  {journal} {\bibinfo  {journal}
  {Physical review letters}\ }\textbf {\bibinfo {volume} {113}},\ \bibinfo
  {pages} {080501} (\bibinfo {year} {2014})}\BibitemShut {NoStop}%
\bibitem [{\citenamefont {Choi}(2015)}]{choi2015dual17}%
  \BibitemOpen
  \bibfield  {author} {\bibinfo {author} {\bibfnamefont {B.-S.}\ \bibnamefont
  {Choi}},\ }\href@noop {} {\bibfield  {journal} {\bibinfo  {journal} {Quantum
  Information Processing}\ }\textbf {\bibinfo {volume} {14}},\ \bibinfo {pages}
  {2775} (\bibinfo {year} {2015})}\BibitemShut {NoStop}%
\bibitem [{\citenamefont {Jochym-O’Connor}\ and\ \citenamefont
  {Laflamme}(2014)}]{jochym2014using18}%
  \BibitemOpen
  \bibfield  {author} {\bibinfo {author} {\bibfnamefont {T.}~\bibnamefont
  {Jochym-O’Connor}}\ and\ \bibinfo {author} {\bibfnamefont {R.}~\bibnamefont
  {Laflamme}},\ }\href@noop {} {\bibfield  {journal} {\bibinfo  {journal}
  {Physical review letters}\ }\textbf {\bibinfo {volume} {112}},\ \bibinfo
  {pages} {010505} (\bibinfo {year} {2014})}\BibitemShut {NoStop}%
\bibitem [{\citenamefont {Yoder}\ \emph {et~al.}(2016)\citenamefont {Yoder},
  \citenamefont {Takagi},\ and\ \citenamefont {Chuang}}]{yoder2016universal25}%
  \BibitemOpen
  \bibfield  {author} {\bibinfo {author} {\bibfnamefont {T.~J.}\ \bibnamefont
  {Yoder}}, \bibinfo {author} {\bibfnamefont {R.}~\bibnamefont {Takagi}}, \
  and\ \bibinfo {author} {\bibfnamefont {I.~L.}\ \bibnamefont {Chuang}},\
  }\href@noop {} {\bibfield  {journal} {\bibinfo  {journal} {Physical Review
  X}\ }\textbf {\bibinfo {volume} {6}},\ \bibinfo {pages} {031039} (\bibinfo
  {year} {2016})}\BibitemShut {NoStop}%
\bibitem [{\citenamefont {Gottesman}(1997)}]{gottesman1997stabilizer22}%
  \BibitemOpen
  \bibfield  {author} {\bibinfo {author} {\bibfnamefont {D.}~\bibnamefont
  {Gottesman}},\ }\href@noop {} {\bibfield  {journal} {\bibinfo  {journal}
  {arXiv preprint quant-ph/9705052}\ } (\bibinfo {year} {1997})}\BibitemShut
  {NoStop}%
\bibitem [{\citenamefont {Gottesman}(2006)}]{gottesman2006quantum24}%
  \BibitemOpen
  \bibfield  {author} {\bibinfo {author} {\bibfnamefont {D.}~\bibnamefont
  {Gottesman}},\ }\href@noop {} {\bibfield  {journal} {\bibinfo  {journal}
  {Quantum Information Processing: From Theory to Experiment}\ }\textbf
  {\bibinfo {volume} {199}},\ \bibinfo {pages} {159} (\bibinfo {year}
  {2006})}\BibitemShut {NoStop}%
\bibitem [{\citenamefont {Chamberland}\ \emph {et~al.}(2016)\citenamefont
  {Chamberland}, \citenamefont {Jochym-O'Connor},\ and\ \citenamefont
  {Laflamme}}]{chamberland2016architectural26}%
  \BibitemOpen
  \bibfield  {author} {\bibinfo {author} {\bibfnamefont {C.}~\bibnamefont
  {Chamberland}}, \bibinfo {author} {\bibfnamefont {T.}~\bibnamefont
  {Jochym-O'Connor}}, \ and\ \bibinfo {author} {\bibfnamefont {R.}~\bibnamefont
  {Laflamme}},\ }\href@noop {} {\bibfield  {journal} {\bibinfo  {journal}
  {arXiv preprint arXiv:1609.07497}\ } (\bibinfo {year} {2016})}\BibitemShut
  {NoStop}%
\end{thebibliography}%

\end{document}